\documentclass[final,5p,times,twocolumn,authoryear]{elsarticle}

\usepackage{amssymb}
\usepackage{amsmath}
\usepackage{lipsum}
\usepackage{xspace}
\usepackage{makecell}
\usepackage{tabularray}
\usepackage{booktabs}
\usepackage{multirow}
\usepackage{graphicx}
\usepackage{algorithm}
\usepackage{algpseudocode}
\usepackage{subcaption}
\usepackage[normalem]{ulem}
\usepackage{tikz}
\usepackage{xcolor}
\usepackage{balance}
\usepackage{hyperref}
\usepackage{xurl}
\newcommand*\circled[1]{\tikz[baseline=(char.base)]{
            \node[shape=circle,fill,inner sep=1pt] (char) {\textcolor{white}{#1}};}}
\renewcommand{\cite}[1]{\citep{#1}}

\useunder{\uline}{\ul}{}

\makeatletter
\newcommand\footnoteref[1]
{\protected@xdef\@thefnmark{\ref{#1}}\@footnotemark}
\makeatother

\newcommand{\quotes}[1]{``#1''}
\def\name{BotHash\xspace}

\def\btype{$\mathbb{B}^3_{\textit{type}}$\xspace}
\def\bcontent{$\mathbb{B}^5_{\textit{content}}$\xspace}
\def\btemporal{$\mathbb{B}^9_{\textit{temporal}}$\xspace}

\journal{ }

\begin{document}

\begin{frontmatter}

\title{\name: Efficient and Training-Free Bot Detection Through Approximate Nearest Neighbor}

\author[first]{Edoardo Di Paolo}
\ead{dipaolo@di.uniroma1.it}
\affiliation[first]{organization={Department of Computer Science, Sapienza University of Rome},
            postcode={00185},
            country={Italy}}

\author[first]{Fabio De Gaspari}

\author[first]{Angelo Spognardi}

\begin{abstract}
  Online Social Networks (OSNs) are a cornerstone in modern society, serving as platforms for diverse content consumption by millions of users each day. However, the challenge of ensuring the accuracy of information shared on these platforms remains significant, especially with the widespread dissemination of disinformation. Social bots ---automated accounts designed to mimic human behavior, frequently spreading misinformation--- represent one of the critical problems of OSNs. The advent of Large Language Models (LLMs) has further complicated bot behaviors, making detection increasingly difficult.
  This paper presents \emph{\name}, an innovative, training-free approach to social bot detection. \name leverages a simplified user representation that enables approximate nearest-neighbor search to detect bots, avoiding the complexities of Deep-Learning model training and large dataset creation.
  We demonstrate that \name effectively differentiates between human and bot accounts, even when state-of-the-art LLMs are employed to generate posts' content. \name offers several advantages over existing methods, including its independence from a training phase, robust performance with minimal ground-truth data, and early detection capabilities, showing promising results across various datasets. 
\end{abstract}


\end{frontmatter}

\section{Introduction}
\label{sec:introduction}
Over the past $20$ years, Online Social Networks (OSNs) have become a cornerstone of modern society. These platforms connect hundreds of millions of individuals every day ---or even billions\footnote{In Q1 2024, Meta reported 3.24 billion active users across their family of platforms.}--- who engage with different types of content, such as news, videos, and images. One of the major strengths of OSNs is the real-time sharing of information, which removes barriers to participation in public discourse. However, this strength also highlights a significant contemporary issue: the spread of misinformation by individuals who are often coordinated in their actions. Central to this issue are social bots, automated accounts designed to mimic the behavior of genuine accounts with the purpose of sharing misinformation or, more generally, harmful content. Estimating the bots' population in OSNs is challenging, but recent studies estimate it could be between 8 and 18\%~\cite{fukuda2022estimating} on X, formerly Twitter.

Recent examples of social bots influencing public discourse include misinformation campaigns during various state elections and the COVID-19 pandemic. Researchers examined OSN users' behavior during elections and uncovered coordinated bot efforts to sway voter decisions by spreading fake news~\cite{bessi2016social,arnaudo2017computational,yang2019bot,de2023twitter}. Similarly, during the COVID-19 pandemic, there were numerous cases of misinformation being spread, which led to shifts in public opinion and encouraged dangerous behaviors, directly causing loss of human life~\cite{covid19_assessing,covid19_social_bots_role}. Equally important are the recent advancements in the field of artificial intelligence, particularly generative AI. The development and widespread availability of Large Language Models (LLMs) have enabled bots to generate text that closely mimics genuine users, making them more credible, significantly harder to detect, and allowing them to integrate more seamlessly into online communities~\cite{feng2024does,li2023you}.
For over a decade, the scientific community has been actively developing various techniques to counter this threat. Many of the most effective detection methods today leverage Deep Neural Networks (DNNs), which excel at learning complex, non-linear relationships between input features to identify malicious activity~\cite{hayawi2022deeprobot,yang2023rosgas,zhou2023detecting}. These models are typically trained on large datasets collected via OSNs' public APIs or directly scraped from their web interfaces. However, partly in response to concerns about the use of such data for training DNNs~\cite{twitter_llm}, many OSNs are implementing measures to significantly restrict data availability and curtail scraping activities~\cite{xapiclosed,scraping_lawsuit}. These restrictions severely hinder the ability to adequately train DNN models, underscoring the need for detection techniques that are less reliant on large volumes of data.

This paper proposes \textit{\name}, a novel, training-free framework for social bot detection. In contrast to many recent approaches that rely on machine learning, \name leverages streamlined user representation that enables approximate nearest-neighbor search for bot identification. \name applies the concept of \textit{Digital DNA}~\cite{cresci2016dna} to encode each OSN user as a sequence of characters based on their activity on the platform. We leverage this simplified encoding to identify similarities between unknown OSN users and a labeled ground truth by integrating \textit{MinHashing}~\cite{broder1997resemblance} and \textit{Locality Sensitive Hashing}~\cite{indyk1998approximate} to classify accounts as either human or bots. Given its prominence and the widespread research interest it attracts, our work focuses on the social network X. However, \name's approach is adaptable and is not limited to this platform alone. 

Our main contributions are summarized as follows:
\begin{itemize}
    \item We present \name, a new, training-free bot detector that does not rely on machine learning or large labeled datasets. \name leverages a simplified OSN user representation and approximate nearest-neighbor search to identify bots.
    \item We propose a new alphabet called \btemporal to represent OSN users, which is designed to capture the temporal features in accounts' behaviors.
    \item We design a new Digital DNA approach, called Multiple-Digital DNA, which leverages multiple alphabet representations to accurately characterize OSN user activity.
    \item We extensively evaluate and compare \name to existing s.o.t.a. techniques across a broad range of datasets from the X platform, showcasing the effectiveness of our approach even against the most recent LLM-based bots.
    \item We demonstrate that our approach generalizes well with only a limited amount of labeled ground-truth data available. Moreover, we show that \name is effective with a small number of tweets per user (as low as 20), facilitating early bot detection.

\end{itemize}

The structure of the paper is as follows: in Section~\ref{sec:related_works}, we present the most relevant literature on the social bot detection task; in Section~\ref{sec:methodology}, we present \name's methodology, discussing the foundations of the approach; in Section~\ref{sec:setup} we present the datasets used in this work and the hardware setup, and in Section~\ref{sec:evaluation} we present \name's evaluation. In Section~\ref{sec:discussion}, we discuss the results and limitations of our approach. Finally, we present our conclusions in Section~\ref{sec:conclusions}.

\section{Related Work}
\label{sec:related_works}
This section discusses a wide range of social bot detection approaches proposed in the literature and compares them with our proposal.

\textbf{Feature-based approaches.}  Several works rely on feature-based approaches for account classification. Yardi et al.~\cite{yardi2010detecting} identify spam content using properties such as username and URL patterns in the tweets. Later, Yang et al.~\cite{yang2013empirical} empirically identify additional sets of features to effectively detect X spammers, reporting state-of-the-art performance on two closed-source datasets. Botometer~\cite{yang2022botometer}, one of the most prominent bot-detection tools, leverages supervised machine learning techniques on more than $1000$ features. However, Botometer was recently shown to fail to detect complex LLM-based bots~\cite{yang2023anatomy}. Efthimion et al.~\cite{efthimion2018supervised} proposed an ML-based approach using usernames' length, temporal patterns, followers/friend ratio, sentiment expression, and re-posting rate for bot classification.

\textbf{Machine Learning-based approaches.}
These approaches leverage Machine Learning (ML) and Deep Learning (DL) algorithms to autonomously learn feature correlations and improve detection performance. Hayawi et al.~\cite{hayawi2022deeprobot} propose a deep neural network, \quotes{DeepProBot}, that exploits user profile data. They used a Long Short-Term Memory (LSTM) network to classify the accounts.
Kudugunta et al.~\cite{kudugunta2018deep} propose an LSTM and Random Forest architectures based on tweets' content and users' metadata. Generative Adversarial Networks (GANs) are used in \quotes{GanBot}, by Najari et al.~\cite{najari2022ganbot}. The authors used a GAN to leak more information about bot accounts, achieving an F1-score of $0.958$ on the Cresci--17 dataset~\cite{cresci2017paradigm}. Mohammad et al.~\cite{mohammad2019bot} employ an architecture based on Convolutional Neural Networks (CNNs) to generate embeddings of posts, which are then used to classify users.

Graph Neural Networks (GNNs) have also been applied to bot detection. Yang et al.~\cite{yang2023rosgas} introduce \quotes{RoSGAS}, a GNN that exploits the connections (e.g., followers/friends relationships) and the features of individual accounts. Feng et al.~\cite{feng2021botrgcn} propose \quotes{BotRGCN}, a GNN architecture that employs user semantic information captured from users' tweets, achieving an F1-score of $0.87$ on the Twibot-20 dataset~\cite{feng2021twibot}. \quotes{DCGNN}~\cite{lyu2023dcgnn}, Dual-Channel Graph Neural Network, focuses on the \quotes{activity-burst} phenomena. The evaluation on the Twibot-22 dataset~\cite{feng2022twibot} showed an F1-score of $0.47$. Finally, Zhou et al.~\cite{zhou2023detecting} present another approach based on GNNs with a contrastive loss. Their framework is based on offline training and online detection, achieving an F1-score of $0.72$ on the Twibot-22 dataset~\cite{feng2022twibot}.

\textbf{Digital-DNA based approaches.} The concept of Digital DNA has been utilized by a variety of approaches. Introduced by Cresci et al. in ~\cite{cresci2016dna,cresci2017paradigm}, the authors observed that bots typically share common sequences of actions. In their original works, they use the longest common DNA substring to classify accounts. Pasricha et al.~\cite{pasricha2019detecting} coupled Digital DNA with lossless compression algorithms and classified accounts using logistic regression. Gilmary et al.~\cite{gilmary2023entropy} propose a framework based on Digital DNA leveraging sequence entropy to classify accounts. Di Paolo et al.~\cite{dipaolo2023dna} use CNNs to transform DNA sequences into images. Chawla et al.~\cite{chawla2023hybrid} used BERT~\cite{DBLP:journals/corr/abs-1810-04805} to extract sentiment information from the tweets and proposed a new alphabet for Digital DNA. Finally, Allegrini et al.~\cite{allegrini2024proposal,allegrini2024deciphering} exploited algorithms for biological DNA to measure the similarity between users' sequences and then they clustered users in macro species based on the similarities.

\textbf{Comparison with current literature.} As of today, most approaches for social bot detection rely on machine or deep learning models. These models require extensive training data, high computational capabilities, and a significant amount of time~\cite{touvron2023llama}. Moreover, machine learning approaches are vulnerable to a variety of adversarial attacks that allow to easily subvert classification results~\cite{de2024have,de2022evading}. Finally, with OSNs recently taking steps to limit data availability~\cite{xapiclosed,scraping_lawsuit}, obtaining large and up-to-date dataset has become challenging. 
\name addresses these challenges by providing a lightweight solution that ensures effective classification with minimal time and memory requirements. Additionally, unlike existing state-of-the-art approaches, \name eliminates the need for training or large datasets, enabling accurate bot detection with as few as 20 tweets per user.
\section{Methodology}
\label{sec:methodology}
\begin{figure*}[t]
    \centering
    \includegraphics[width=\textwidth]{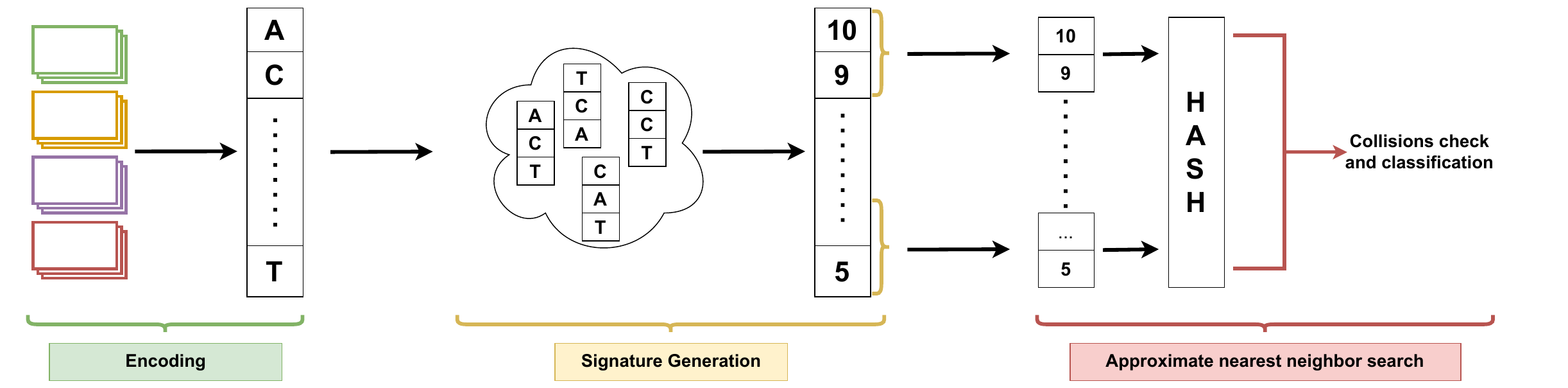}
    \caption{High-Level overview of the \name classification pipeline. \name comprises three main steps: account encoding, signature generation from the encoded vector, and approximate nearest neighbor search on the signature vectors.}
    \label{fig:overview}
\end{figure*}

This section presents the methodology behind \name, including a detailed overview of the simplified user representation and the approximate nearest neighbor search algorithm employed.

\subsection{High-Level Overview}
\label{subsec:overview}
At a high level, \name can be divided into three distinct steps, as highlighted in Figure~\ref{fig:overview}: (1) user activity encoding, (2) user signature generation, and (3) approximate nearest neighbor search. The user encoding step (1) generates a vector representation for each user, processing their post history in the OSN platform. Vector encoding is carried out based on the high-level characteristics of the user posts, while the encoding details depend on the social platform considered.
After obtaining a vector encoding for each user, we tokenize the vector with fixed-length tokens and obtain a set representation. This set representation is further processed in step (2) through MinHashing to generate a unique signature for each user. Finally, in step (3) we perform an approximate nearest neighbor search for the MinHash signature against a labeled ground truth using Locality Sensitive Hashing to identify bots.

\name's approach offers several advantages over existing techniques. The first key advantage is its speed and efficiency. Constructing the ground truth for classification involves a straightforward encoding procedure, shingling, and MinHashing to generate a signature for each user. As detailed in Section~\ref{sec:evaluation}, our experiments demonstrate that this process requires mere seconds, even when dealing with large datasets, in contrast to the hours ---or even days~\cite{touvron2023llama}--- needed by large DNNs. The classification of unknown users is also remarkably fast. The approximate nearest neighbor search employed by \name relies on hash function collisions, which can be found in constant time and is easily parallelizable. Lastly, since each user in the ground truth is represented as a single signature, the memory footprint of \name is limited, as confirmed in our experimental evaluation.

The second significant advantage is that \name does not necessitate a substantial ground truth dataset. We illustrate in our experimental evaluation that the hash-based approximate nearest neighbor search remains effective even with small subsets of the dataset used as ground truth. This characteristic renders \name applicable in scenarios where machine learning-based solutions are impractical due to a scarcity of labeled data.

The third significant advantage of \name lies in its simplified user encoding, which enables robust performance even with limited user data (i.e., a small number of posts), thereby facilitating early bot detection. Our evaluation demonstrates that when the maximum number of posts per user is limited, \name remains effective with as few as 20 posts.

\subsection{User Encoding}
\label{sec:encoding}
The foundation of \name primarily rests on the encoding used to represent a user activity, which yields signatures that are both simple and descriptive enough to embody meaningful features of the user.
These signatures are based on the encoding technique known as \textit{Digital DNA} (D-DNA), initially introduced in ~\cite{cresci2016dna}. D-DNA encoding is inspired by the structure of biological DNA, which is comprised of a combination of four different nucleotides. The idea of D-DNA lies in representing the behavior of an account using an alphabet (corresponding to the nucleotides of biologial DNA), consisting of multiple elements representing different properties and attributes of the account. In this work, we extend the concept of D-DNA by proposing \textit{Multiple Digital DNA} (MD-DNA), which leverages multiple alphabet representations to accurately characterize OSN user activity. Additionally, we introduce a new temporal alphabet that facilitates the capture of temporal patterns in users' behaviors, offering a richer and more detailed encoding.

\subsubsection{Digital DNA} During the pre-processing phase, D-DNA allows us to encode each user in the dataset as a vector defined as:
\begin{equation}
    \label{eq:dnasequence}
    s = [\sigma_1, \sigma_2, \ldots, \sigma_n], \, \sigma_i \in \mathbb{B} \, \forall i = 1 \ldots n.
\end{equation}

where $\mathbb{B}$ is an alphabet that captures specific features of the users' behaviors and each $\sigma_i$ represents the encoding of the $i-$th tweet posted by the user. 
We consider two existing alphabets---\btype~\cite{cresci2016dna}  and \bcontent~\cite{chawla2023hybrid}---and a new alphabet \btemporal. The \btype alphabet (\autoref{eq:alphabetb3type}) consists of three symbols encoding tweet types, differentiating between normal tweets, retweets, and replies. On the other hand, the \bcontent alphabet (\autoref{eq:alphabetb5content}) focuses on tweet content, identifying the presence of URLs, hashtags, mentions or mixed entities. 

\begin{equation}
    \label{eq:alphabetb3type}
    \mathbb{B}^{3}_{\textit{type}} =
    \begin{Bmatrix}
    A \Rightarrow \text{plain tweet} \\ 
    C \Rightarrow \text{retweet} \\
    T \Rightarrow \text{reply} \\ 
    \end{Bmatrix}
\end{equation}

\begin{equation}
    \label{eq:alphabetb5content}
    \mathbb{B}^{5}_{\textit{content}} =
    \begin{Bmatrix}
    X \Rightarrow \text{tweet contains mixed entities}\footnotemark \\ 
    U \Rightarrow \text{tweet contains just URLs} \\
    H \Rightarrow \text{tweet contains just hashtags} \\ 
    M \Rightarrow \text{tweet contains just mentions} \\
    N \Rightarrow \text{all other cases}
    \end{Bmatrix}
\end{equation}\footnotetext{For mixed entities we refer to URLs, hashtags, and mentions.}

Unlike \btype and \bcontent, the \btemporal alphabet captures temporal dynamics in users behaviors byencoding time intervals ($TD$ in \autoref{eq:alphabetb9temporal}) between consecutive tweets. This addition enhances the D-DNA framework by incorporating temporal aspects of user behavior---an element not previously utilized in this context. Prior work has shown that such temporal characteristics can improve detection of bot accounts~\cite{echeverria2019discovery,kirn2021bayesian,bellutta2023investigating}. Appendix~\ref{sec:implementation_details} provides the detailed definition of each alphabet.

\begin{equation}
    \label{eq:alphabetb9temporal}
    \mathbb{B}^{9}_{\textit{temporal}} =
    \begin{Bmatrix}
    B \Rightarrow TD \leq \text{1 hour} \\ 
    D \Rightarrow \text{1 hour} < TD \leq \text{5 hours} \\
    E \Rightarrow \text{5 hours} < TD \leq \text{10 hours} \\
    F \Rightarrow \text{10 hours} < TD \leq \text{15 hours} \\
    G \Rightarrow \text{15 hours} < TD \leq \text{20 hours} \\
    J \Rightarrow \text{20 hours} < TD \leq \text{1 day} \\
    K \Rightarrow \text{1 day} < TD \leq \text{1 week} \\
    I \Rightarrow \text{1 week} < TD \leq \text{1 month} \\
    L \Rightarrow TD > \text{1 month} \\
    \end{Bmatrix}
\end{equation}

While the apparent simplicity of D-DNA encoding may appear limiting, its combination with \name signature generation provides surprising flexibility and representational power. The MinHashing technique used for signature generation highly depends on the cardinality of the alphabet used for encoding, as discussed in Section~\ref{subsec:minhash}. Consequently, while more complex encodings offer more detailed representations, they do not always lead to better signatures and classification. As we show later, alphabet selection largely depends on dataset complexity, with richer alphabets that account for more complex user behaviors proving more effective for complex datasets.

\subsubsection{Multiple D-DNA} We further extend the concept of D-DNA to MD-DNA. In the literature, alphabets have been typically used in insolation~\cite{cresci2016dna,dipaolo2023dna}, limiting the encoding expressiveness. \name overcomes this limitation by simultaneously leveraging multiple alphabets. MD-DNA generalizes the definition in \autoref{eq:dnasequence} by incorporating $N$ distinct alphabets $\mathbb{B}_1, \mathbb{B}_2, \ldots, \mathbb{B}_N$. For each user's tweet, MD-DNA assigns a vector of symbols, one for each alphabet, as follows:
\begin{equation} 
\label{eq:mddnasequence} 
s = [\vec{\sigma}_1, \vec{\sigma}_2, \ldots, \vec{\sigma}_n], \quad \vec{\sigma}_i = (\sigma_i^{(1)}, \sigma_i^{(2)}, \ldots, \sigma_i^{(N)}).
\end{equation}

where
\begin{equation*}
 \sigma_i^{(j)} \in \mathbb{B}_j, \forall j =1, \ldots, N.
\end{equation*}

In other words, each element of the sequence $s$ is represented by a vector $\vec{\sigma}_i$ that simultaneously captures information from $N$ alphabets. This approach allows for the integration of heterogeneous information, such as tweet types (\btype), tweet contents (\bcontent), and temporal characteristics (\btemporal) into a single encoding.
To better understand the MD-DNA, let us consider a simple example using the \btype and \bcontent alphabets. Suppose we have an account with just three tweets: for the \btype alphabet we obtain a sequence like \texttt{ACT}, while for the \bcontent \texttt{HUM}. Thus, the MD-DNA sequence for this example would be \texttt{AHCUTM}, in which each vector $\vec{\sigma_i}$ defines a pair of symbols.

\subsection{MinHashes Generation}
\label{subsec:minhash}
The symbols sequence $s$ used to encode users' activity is then processed with the \textit{MinHashing} technique to generate a signature that efficiently captures the similarity between different users. MinHashing was initially developed to generate signatures of documents and efficiently find similarities between them~\cite{broder1997resemblance,broder2000identifying}, but can be generally employed to find similarities between any two arbitrary sets. The input of the MinHashing process is a set of tokens (\textit{shingles}) obtained from the encoded user vector of step (1) applying the process known as \textit{shingling}. Shingling involves dividing the encoded vector in shingles using a sliding window of fixed size, ultimately resulting in a set of shingles. The size of the sliding window is an hyperparameter of the algorithm and can be tuned based on the specific dataset and alphabets used. In the remainder of this paper, we refer to this hyperparameter as \emph{k-shingle}. More formally, let $\mathbb{B}$ be an alphabet and $\mathcal{U}$ the vector representation of user $U$ defined as

\begin{equation*}
    \label{eq:dnasequence_example}
    \mathcal{U} = [\sigma_1, \ldots, \sigma_n], \, \sigma_i \in \mathbb{B} \, \forall i = 1 \ldots n
\end{equation*}

then, the set representation of user $\mathcal{U}$ after the shingling procedure with k-shingle parameter $k$ is:
\begin{equation*}
    \label{eq:setshingle_example}
    \mathcal{F} = \left\{ [\sigma_i, \ldots, \sigma_{i+k}] : 1 \leq i \leq n-k\right\} 
\end{equation*}

Starting from the set representation of each user, we rely on MinHashing to generate a short, descriptive signature that preserves the similarity between the sets. The limited size of the signature compared to the original size of the set $\mathcal{F}$ is a key aspect for the efficiency and speed of \name's classification. As its core, MinHashing involves creating a signature for each set by hashing its elements based on a permutation of all the possible elements of a set, called \textit{Universe}. The size of the Universe directly depends on the cardinality of the alphabet used in step (1), and the k-shingle hyperparameter. The number of permutations used for the MinHashing directly influences the final generated signature, and typically a larger number of permutations leads to more accurate results, but also increases computational complexity. A trade-off between speed and accuracy is therefore required.

\subsection{Approximate Nearest Neighbor Search}
\label{subsec:lsh}
The final step in the \name classification pipeline involves matching the signature of the query user (i.e., the unknown user to be classified) to other similar user signatures in the ground truth. Based on the number of similar human and bot users and their distance (the Jaccard similarity, \autoref{eq:jaccard_similarity}), the query user is classified either as human or bot. In order to be effective in this context, an ideal nearest-neighbor search algorithm should have two key properties: (1) fast and efficient and (2) local sensitivity. The first property is self-explanatory. In contrast to commonly used DNN approaches, \name is built to be fast, scalable, and to generalize with few labeled data. Consequently, excessively complex or machine-learning-based nearest-neighbor search techniques are not applicable. The second key property, local sensitivity, derives from the expected behaviors of bots in OSNs. Previous research on X bot detection emphasized that, although the overall behavior of bots on the platform may exhibit significant variation, there exists a set of sequences of actions that are consistently observed across bots~\cite{cresci2016dna}. Consequently, while the overall activity of different bots on the OSN platform may differ, the identification and correlation of common sequences of actions enables effective classification. Hence, a nearest neighbor search algorithm that is sensitive to local commonalities within the vectors generated by the signature generation step is essential.

Given the above-mentioned requirements, \name utilizes \textit{Locality Sensitive Hashing} (LSH)~\cite{indyk1998approximate} to classify users as either bots or humans. Unlike conventional hashing functions, LSH is specifically designed to maximize collisions between similar input vectors. The primary characteristic of LSH enabling this behavior is its partitioning of the input vector into multiple sub-vectors, each of which undergoes independent hashing into a set of buckets. Any collision with other samples within these buckets indicates a potential similarity between the input and collided samples. Essentially, LSH is designed to facilitate the clustering of similar entities into the same buckets with a high likelihood while ensuring the placement of dissimilar entities in separate buckets. Because LSH relies on simple hash functions, the computation is extremely fast, allowing for the identification of similar candidates (i.e., users with similar signatures) in constant time (1). Furthermore, the independent hashing of individual sub-vectors by LSH ensures that the neighbor search is locally sensitive (2), facilitating the identification of common behavioral subsequences within a user's overall OSN activity.

\begin{equation}
    \label{eq:jaccard_similarity}
    J(A, B) = \frac{|A \cap B|}{|A \cup B|}
\end{equation}

The number of sub-vectors used during the LSH process is implicitly specified through a Jaccard similarity hyperparameter, which approximately indicates the minimum Jaccard similarity between two signatures necessary for them to collide in at least one bucket. In \name, $A$ and $B$ in \autoref{eq:jaccard_similarity} represent the DNA signature of two accounts.
\begin{equation}
    \label{eq:classification}
    \hat{y}_u = \begin{cases} 
\text{bot} & \text{if} \quad \sum_{v \in N_u} y_v > \frac{|N_u|}{2} \\
\text{human} & \text{otherwise}
\end{cases}
\end{equation}
The classification of an individual query sample is carried out using a majority voting rule based on the neighbors' labels, as shown in \autoref{eq:classification}; here $\hat{y}_u$ represents the predicted class label of the query user $u$, which is either bot or human. The sum $\sum_{v \in N_u} y_v$ represents the sum of the labels of the nearest neighbors $N_u$. The query user is classified as a bot if more than half of its neighbors are labeled as bots, and as human otherwise.
\section{Experimental Setup}
\label{sec:setup}
This section presents the dataset used in our evaluation and the experimental setup used to run the experiments.

\subsection{Datasets}
\label{subsec:datasets}

To assess the effectiveness and potential of our approach, we selected several publicly available datasets from the literature. Notably, due to recent restrictions on X's API, acquiring new data has become increasingly challenging. In our dataset selection, we aimed to include both recent, up-to-date datasets and widely-used benchmark datasets in the field. All raw data used in our evaluation is available at the \quotes{bot repository}\footnote{\href{https://botometer.osome.iu.edu/bot-repository/}{https://botometer.osome.iu.edu/bot-repository/}}.

\begin{table}
\caption{Datasets used in our evaluation. The \quotes{All} dataset is the union of all the datasets.}
\label{tab:datasets_table}
\centering
\resizebox{.8\columnwidth}{!}{%
\begin{tabular}{cccc}
\hline
\textbf{Dataset} & \textbf{\begin{tabular}[c]{@{}c@{}}Human \\ accounts\end{tabular}} & \textbf{\begin{tabular}[c]{@{}c@{}}Bot \\ accounts\end{tabular}} & \textbf{\begin{tabular}[c]{@{}c@{}}Total \\ accounts\end{tabular}} \\ \hline
Cresci-2015~\cite{cresci2015fame} & 1946 & 3202 & 5148 \\ \hline
Cresci-2017~\cite{cresci2017social} & 1083 & 9114 & 10197 \\ \hline
Cresci--2018~\cite{cresci2018fake} & 7479 & 18509 & 25988 \\ \hline
Twibot--22~\cite{feng2022twibot} & 95153 & 82463 & 177616 \\ \hline
Fox-8~\cite{yang2023anatomy} & 1139 & 1140 & 2279 \\ \hline
All & 106734 & 111219 & 217953 \\ \hline
\end{tabular}%
}
\end{table}

\autoref{tab:datasets_table} provides an overview of the datasets' sizes; however, we underline that during the preprocessing phase, some users may be removed based on the \textit{k-shingle} parameter. Specifically, if an account has fewer tweets than the \textit{k-shingle} number (typically, less than 10), it will be excluded during the evaluation phase. Further details are provided below.

\textbf{Cresci--2015.} The Cresci--2015 was originally presented by Cresci et al. in~\cite{cresci2015fame}. It includes different categories of human accounts and bot accounts in the fake follower category. The dataset consists of $1946$ human accounts and $3202$ bot accounts, for a total of $5148$ accounts. It is a widely-used benchmark in the field~\cite{feng2022twibot}.

\textbf{Cresci--2017.} This dataset was introduced by Cresci et al. in ~\cite{cresci2017paradigm} in 2017, and it stands as a major benchmark in the social bot detection literature~\cite{feng2021twibot,feng2022twibot}. The dataset encompasses several categories of social bots, including those involved in online political discussions, promoting specific hashtags, and advertising Amazon products and paid apps for mobile devices. The Cresci--2017 dataset is an important reference because it includes both simple spambots that repeatedly tweet the same posts, as well as social spambots that mimic the profiles and actions of normal users~\cite{yang2020scalable}. 
Each user in the dataset has numerous tweets, and no accounts were removed from the dataset in the preprocessing step.
The unbalanced nature of the dataset serves as a good benchmark to evaluate the performance of \name in challenging scenarios where the ground truth is highly unbalanced towards one of the classes.

\textbf{Cresci--2018.} Originally presented in~\cite{cresci2018fake} in 2018, this dataset comprises human and bot accounts engaged in disseminating tweets relating to stocks and financial markets. The dataset focuses on bots using \textit{cashtags} and coordinating their OSN activity to promote low-value stocks for speculative campaigns. Each account consists of numerous tweets, and no accounts were removed during the preprocessing step. Similar to Cresci--17, the dataset is highly unbalanced, with bots representing over two-thirds of the dataset.

\textbf{Twibot--22.} 
Twibot--22 is the largest and most comprehensive dataset used in this study. Originally introduced in~\cite{feng2022twibot} in 2022, this dataset contains one million labeled accounts; approximately $860,000$ are humans, while $140,000$ are bots. Due to the large size of the dataset and resource constraints, existing studies typically utilize only a subset of the data~\cite{zhou2023detecting}. Similarly, we sample a subset of the accounts from the dataset and preprocess them, resulting in a total of $177,616$ accounts, of which $95,153$ are humans and $82,463$ are bots.

\textbf{Fox--8.} Introduced by Yang et al. in 2023~\cite{yang2023anatomy}, the Fox--8 dataset is one of the earliest contributions in the literature featuring bot accounts that generate text through LLMs. The authors identified LLM-powered bots using heuristic methods by examining tweets where the bots inadvertently revealed their nature, often through a standardized message that identified them as language models and mentioned their inability to fulfill certain requests. Yang et al. identified 1,140 LLM-powered bots, primarily used for promoting cryptocurrency and blockchain websites. Fox--8 is a valuable dataset for evaluating \name's ability to detect new-generation social bots that leverage LLMs to generate human-like content. As demonstrated by the authors in their paper~\cite{yang2023anatomy}, samples in the Fox--8 dataset are generally challenging for both social bot detectors and LLM detectors to identify.

\subsection{Hardware Setup}
\label{sec:hardware_setup}
The experiments reported in this study were conducted on a single workstation equipped with an AMD Ryzen 9 7950X 16-Core Processor running at 4.50GHz, 64 GB of RAM, and an RTX 4090 GPU with 24 GB of memory. It is important to note that \name is not fully optimized; for instance, the computations for hash generation are performed on the CPU rather than the GPU. Optimizing the underlying library and code would likely lead to even faster performance.
\section{Evaluation}
\label{sec:evaluation}
This section presents the evaluation of \name. Section~\ref{subsec:bot_analysis} presents detection results for \name, along with a comparison to other state-of-the-art approaches. Section~\ref{sec:resource_usage} analyzes the resource usage of \name and compares it to other state-of-the-art ML and DL techniques. Section~\ref{subsec:generalization_edetection} analyzes the cross-dataset generalization and early detection capabilities of \name, as well as its performance when only limited ground truth data is available.

\subsection{Bot Detection Analysis}
\label{subsec:bot_analysis}

\begin{table*}[]
\centering
\caption{Results obtained using BotHash with a random data split. Specifically, 70\% of the data was used as ground truth set, while 30\% was as test set. For reasons of space, we report only the F1 score and Accuracy. Best results are in bold.}
\label{tab:our-split-results}
\resizebox{\textwidth}{!}{%
\begin{tabular}{@{}ccccccccccccccccccccc@{}}
\toprule
\textbf{Dataset} & \multicolumn{20}{c}{\textbf{Alphabets}} \\ \midrule
 & \multicolumn{2}{c}{\btype} &  & \multicolumn{2}{c}{\bcontent} &  & \multicolumn{2}{c}{\btemporal} &  & \multicolumn{2}{c}{\btype + \bcontent} &  & \multicolumn{2}{c}{\btype + \btemporal} &  & \multicolumn{2}{c}{\bcontent + \btemporal} &  & \multicolumn{2}{c}{\btype + \bcontent + \btemporal} \\ \cmidrule(lr){2-3} \cmidrule(lr){5-6} \cmidrule(lr){8-9} \cmidrule(lr){11-12} \cmidrule(lr){14-15} \cmidrule(lr){17-18} \cmidrule(l){20-21} 
 & F1 & Acc. &  & F1 & Acc. &  & F1 & Acc. &  & F1 & Acc. &  & F1 & Acc. &  & F1 & Acc. &  & F1 & Acc. \\ \cmidrule(lr){2-3} \cmidrule(lr){5-6} \cmidrule(lr){8-9} \cmidrule(lr){11-12} \cmidrule(lr){14-15} \cmidrule(lr){17-18} \cmidrule(l){20-21} 
\textbf{Cresci--15} & 95.84 & 95 &  & 95.57 & 94.59 &  & 93.81 & 92.37 &  & \textbf{96.10} & \textbf{95.42} &  & 94.71 & 93.5 &  & 95.53 & 94.63 &  & 95.90 & 95.07 \\ \midrule
\textbf{Cresci--17} & 98.74 & 97.79 &  & 98.04 & 96.54 &  & 96.69 & 94.29 &  & \textbf{98.78} & \textbf{97.84} &  & 98.51 & 97.35 &  & 97.94 & 96.39 &  & 98.46 & 97.30 \\ \midrule
\textbf{Cresci--18} & \textbf{98.67} & \textbf{97.55} &  & 94.64 & 90.12 &  & 93.05 & 87.36 &  & 97.30 & 94.95 &  & 95.83 & 92.23 &  & 93.68 & 88.44 &  & 95.59 & 91.76 \\ \midrule
\textbf{Twibot-22} & 60.89 & 56.54 &  & 62.91 & 59.48 &  & 61.88 & \textbf{61.68} &  & 62.16 & 60.46 &  & 63.07 & 60.88 &  & 63.68 & 59.87 &  & \textbf{64.25} & 61.65 \\ \midrule
\textbf{Fox-8} & 94.08 & 93.83 &  & 87.89 & 86.95 &  & 94.79 & 94.59 &  & 92.09 & 91.78 &  & \textbf{98.39} & \textbf{98.97} &  & 96.30 & 96.33 &  & 97.89 & 97.80 \\ \bottomrule
\end{tabular}%
}
\end{table*}

In this section, we present the results achieved by \name on the datasets presented in Subsection~\ref{subsec:datasets}. 
To ensure a thorough and fair evaluation, we present two sets of results: (1) ransom-split alphabet comparison and (2) fixed-split s.o.t.a. comparison. 

\subsubsection{Random-Split Alphabet Comparison}
This section presents \name's performance when using different encoding alphabets. Following standard practice, we randomly split each dataset with a 70\% - 30\% ratio between the ground truth set and the test set.

\textbf{Cresci--2015.}
\label{subsubsec:osr-cresci-2015}
\autoref{tab:our-split-results} shows that the highest F1-score, obtained by our framework, is $96.10\%$ using the MD-DNA approach with the \btype and \bcontent alphabets. This result highlights the effectiveness of combining the tweet type and content features in capturing bot and human behaviors.

\textbf{Cresci--2017.}
\label{subsubsec:osr-cresci-2017}
As previously discussed, the Cresci--2017 dataset comprises a diverse mix of social bots, making it a crucial benchmark for evaluating the generalizability of bot detectors. As highlighted in~\autoref{tab:our-split-results}, \name demonstrates remarkable performance on this dataset, achieving exceptionally high accuracy and high F1-score. These results highlight the ability of our approach to recognize complex bot behaviors starting from a simplified alphabet, such as the \btype. Moreover, it highlights how the MD-DNA approach can combine the strengths of multiple alphabets, resulting in stronger classification results.

\textbf{Cresci--2018.}
\label{subsubsec:osr-cresci-stock-2018}
This dataset is known to present a higher challenge for bot detection compared to the Cresci--2017 dataset~\cite{feng2022twibot}, likely due to the nature of the bot accounts considered. However, \name shows remarkable classification performance that is in line with the previous Cresci datasets. Surprisingly, in this test, the MD-DNA approach shows lower performance than the classical D-DNA approach on this dataset; in fact, the simplest encoding sequences alphabet, \btype, provides the highest performance. It is challenging to identify the reasons behind this result. It is likely a combination of the nature of bots included in the dataset, which results in mainly homogeneous content using cashtags, and the low correlation of temporal features with the bot's behavior. 

\textbf{Twibot--22.} 
\label{subsubsec:osr-twibot-22}
The Twibot--22 dataset comprises heterogeneous accounts that exhibit considerably different patterns. It is typically one of the most challenging datasets for bot detection~\cite{feng2022twibot}. As highlighted in \autoref{tab:our-split-results}, \name's performance decreases compared to previous datasets. Interestingly, the temporal alphabet \btemporal is arguably the best single-encoding approach, with performance close to \bcontent. These results suggest that, at a large scale, temporal characteristics of posts are a valuable indicator of bot behavior. Furthermore, MD-DNA encodings show generally superior performance compared to single D-DNA encodings, indicating that a combination of type of posts, content, and temporal characteristics are necessary to fully characterize heterogeneous bot behaviors.

\textbf{Fox-8.}
\label{subsubsec:osr-fox-8}
The Fox--8 dataset is an interesting benchmark for \name, as it consists entirely of LLM-generated tweets, which are designed to closely resemble human-generated content. As illustrated in~\autoref{tab:our-split-results}, \name performs exceptionally well on this dataset, achieving an F1 of $98.39\%$ exploiting the MD-DNA with \btype and \btemporal alphabets. These results underscore the robustness of \name's user encoding and local sensitivity when faced with bots that generate human-like texts. We highlight that the authors of Fox-8 reported that Botometer~\cite{yang2022botometer} consistently fails to detect bot accounts, with an average recall of $4\%$~\cite{yang2023anatomy}. Additionally, the authors evaluated tools specifically designed to detect LLM-generated text and they reported that the best tool considered ---OpenAI's LLM detector--- achieved an F1 score of $84\%$~\cite{yang2023anatomy}, significantly lower than \name's performance.

\textbf{Takeaway.} Overall, the results in Table~\ref{tab:our-split-results} demonstrate \name's ability to reliably distinguish bots and human accounts, even when state-of-the-art LLMs are used to generate post content. This effectiveness is attributed to \name's capability to analyze the similarity of subsequences of user actions. These findings validate our hypothesis that \textit{local sensitivity} is a critical property for effective social bot detection. Furthermore, the results underscore that the MD-DNA approach generally enhances performance, with a key contributing factor being the individual effectiveness of each alphabet. Finally, more complex alphabets are not always inherently superior to simpler ones. This is evident from the performance of \bcontent on the Fox--8 dataset. This outcome is expected, as \bcontent incorporates post content into its encoding, and the bot-generated content in Fox--8, created using LLMs, closely resembles human-generated content.

\subsubsection{Fixed-Split SOTA Comparison.}
\label{sec:sota_comparison}
This section compares \name with other well-known state-of-the-art approaches. To provide a comprehensive analysis, we include both ML-based and DL-based techniques in the evaluation. As ML-based models, we selected three widely-used benchmark papers in the area: Efthimion~\cite{efthimion2018supervised}, Kudugunta~\cite{kudugunta2018deep}, and SGBot~\cite{yang2020scalable}. For the DL-based models, we selected two recent graph-based approaches that show good performance on benchmarks~\cite{feng2021twibot}: BotRGCN~\cite{feng2021botrgcn}, and RGT~\cite{feng2022heterogeneity}. We base our evaluation on the implementations in~\cite{feng2022twibot}. To ensure consistency with existing works, all results in this section are based on the train-test splits provided by the authors of Twibot--22~\cite{feng2022twibot}. As noted in Section~\ref{subsec:datasets}, due to Twibot-22's large size, we randomly sample a subset of the data from the splits provided by the authors for our evaluation. For completeness, we report the hyperparameters used for \name in \autoref{tab:hyperparameters_table}, which were found through grid search.
\autoref{tab:sota-comparison} summarizes our results.

\textbf{Cresci--15.}
As illustrated in~\autoref{tab:sota-comparison}, \name achieves the second-best F1 score on Cresci--15, with a $\sim96\%$ performance. This result highlights \name's ability to classify bot accounts more effectively than much more complex DL-based models~\cite{feng2021botrgcn,feng2022heterogeneity}. Moreover, compared to other models such as~\cite{kudugunta2018deep}, where the recall is significantly lower, \name demonstrates the ability to achieve a much more balanced classification in terms of precision and recall.

\textbf{Cresci--17.}
The Cresci--17 dataset lacks information on user relationships, preventing the evaluation of graph-based models such as BotRGCN and RGT, which rely on this data. Among the remaining techniques, \name achieves superior performance, outperforming existing approaches. The second-best approach, SGBot, achieves an F1-score of $96.22\%$, nearly $3$ percentage points lower than \name, which achieves $99.06\%$. Notably, most existing detectors exhibit a significant imbalance between the ability to effectively recognize bots (recall) and their detection confidence (precision). This imbalance implies that one must either accept a high detection rate with a high false positive rate (i.e., benign users detected as bots) or require low false positive rates at the expense of the detection rate. \name avoids this tradeoff, offering precise bot detection capabilities without compromising the detection rate.

\begin{table*}[]
\centering
\caption{
Comparison with the state of the art. The bold values represent the best results obtained among all the selected models, while the underlined values indicate the second-best results. The `-' symbol denotes that the model is not applicable to the dataset.}
\label{tab:sota-comparison}
\resizebox{\textwidth}{!}{%
\begin{tabular}{@{}ccccccccccccccccccccccccc@{}}
\toprule
\textbf{Model} & \multicolumn{24}{c}{\textbf{Datasets}} \\ \midrule
 & \multicolumn{4}{c}{\textbf{Cresci--15}} &  & \multicolumn{4}{c}{\textbf{Cresci--17}} &  & \multicolumn{4}{c}{\textbf{Cresci--18}} &  & \multicolumn{4}{c}{\textbf{Twibot--22}} &  & \multicolumn{4}{c}{\textbf{Fox--8}} \\ \midrule
\textbf{} & F1 & Acc. & Prec. & Rec. &  & F1 & Acc. & Prec. & Rec. &  & F1 & Acc. & Prec. & Rec. &  & F1 & Acc. & Prec. & Rec. &  & F1 & Acc. & Prec. & Rec. \\ \cmidrule(lr){2-5} \cmidrule(lr){7-10} \cmidrule(lr){12-15} \cmidrule(lr){17-20} \cmidrule(l){22-25} 
\textbf{Efthimion}~\cite{efthimion2018supervised} & 92.39 & 92.63 & 89.77 & {\ul 95.18} &  & 95.94 & 92.91 & 96.24 & {\ul 95.64} &  & 82.14 & 75.68 & 81.05 & 83.25 &  & 56.15 & 55.16 & 83.34 & 42.33 &  & 88.81 & 85.96 & 81.93 & 96.94 \\
\textbf{Kudugunta}~\cite{kudugunta2018deep} & 90.15 & {\ul 97.56} & \textbf{100} & 82.08 &  & 92.43 & {\ul 95.14} & 99.19 & 86.53 &  & 83.46 & \textbf{95.92} & 78.65 & 88.94 &  & {\ul 73.29} & \textbf{83.9} & 60.83 & \textbf{92.18} &  & 97.59 & 97.28 & {\ul 99.05} & 96.18 \\
\textbf{BotRGCN}~\cite{feng2021botrgcn} & 68.70 & 56.37 & 82.44 & 58.89 &  & - & - & - & - &  & - & - & - & - &  & 71.70 & 66.81 & {\ul 85} & 62 &  & - & - & - & - \\
\textbf{RGT}~\cite{feng2022heterogeneity} & 70.18 & 58.35 & 83.98 & 60.27 &  & - & - & - & - &  & - & - & - & - &  & 71.75 & {\ul 66.98} & \textbf{85.46} & 61.82 &  & - & - & - & - \\
\textbf{SGBot}~\cite{yang2020scalable} & \textbf{98.18} & \textbf{98.3} & {\ul 98.78} & \textbf{97.59} &  & {\ul 96.22} & 93.57 & {\ul 99.2} & 93.4 &  & {\ul 88.19} & 83.89 & \textbf{86.84} & {\ul 89.59} &  & 69.43 & 63.16 & 80.84 & 60.83 &  & {\ul 98.08} & {\ul 97.80} & 98.46 & {\ul 97.77} \\ \midrule
\textbf{BotHash} & {\ul 95.94} & 95.11 & 96.29 & 94.97 &  & \textbf{99.06} & \textbf{98.34} & \textbf{99.46} & \textbf{98.67} &  & \textbf{91.03} & {\ul 86.21} & {\ul 83.99} & \textbf{99.15} &  & \textbf{75.83} & 64.40 & 71.10 & {\ul 81.25} &  & \textbf{98.39} & \textbf{98.97} & \textbf{99.69} & \textbf{98.19} \\ \bottomrule
\end{tabular}%
}
\end{table*}
\begin{table*}[]
\centering
\caption{Time and memory usage requirements for each considered technique. Preprocessing and training time are presented in seconds. Memory usage represents the MBs used by the approach in the RAM. The symbol `-' denotes that the model is not applicable to the dataset.}
\label{tab:timings-table}
\resizebox{\textwidth}{!}{%
\begin{tabular}{cccccccccccccccc}
\hline
\textbf{Model} & \multicolumn{15}{c}{\textbf{Dataset}} \\ \hline
 & \multicolumn{3}{c}{\textbf{Cresci-2015}} &  & \multicolumn{3}{c}{\textbf{Cresci-2017}} &  & \multicolumn{3}{c}{\textbf{Cresci-2018}} &  & \multicolumn{3}{c}{\textbf{Twibot-22}} \\ \midrule 
 & Preprocess & Training & Mem. usage &  & Preprocess  & Training  & Mem. usage &  & Preprocess  & Training  & Mem. usage &  & Preprocess  & Training  & Mem. usage \\ \cline{2-4} \cline{6-8} \cline{10-12} \cline{14-16} 
\textbf{Efthimion}~\cite{efthimion2018supervised}& 5.1 & 0.03 & 2185 &  & 13.49 & 0.1 & 6619 &  & 0.62 & 0.04 & 139 &  & 11.06 & 93.95 & 4806 \\
\textbf{Kudugunta}~\cite{kudugunta2018deep} & 0.32 & 0.19 & 1229 &  & 0.78 & 0.29 & 2697 &  & 0.01 & 0.15 & 121 &  & 0.67 & 9.46 & 3087 \\
\textbf{BotRGCN}~\cite{feng2021botrgcn} & 818.15 & 0.53 & 939 &  & - & - & - &  & - & - & - &  & 27900 & 2.12 & 1426 \\
\textbf{RGT}~\cite{feng2022heterogeneity} & 818.15 & 129.43 & 21600 &  & - & - & - &  & - & - & - &  & 27900 & 1656.9 & 22530 \\
\textbf{SGBot}~\cite{yang2020scalable} & 0.68 & 0.22 & 98 &  & 0.49 & 0.39 & 106 &  & 0.48 & 0.21 & 100 &  & 36.22 & 20.69 & 624 \\
\midrule
\textbf{\name} & 1.54 & 1.17 & 128 &  & 3.38 & 1.43 & 190 &  & 0.42 & 1.84 & 180 &  & 9.63 & 10.48 & 1933 \\ \hline
\end{tabular}%
}
\end{table*}

\textbf{Cresci--18.}
As discussed in Section~\ref{subsubsec:osr-cresci-stock-2018}, this dataset is more challenging compared to the previous ones, such as Cresci--15 and Cresci--17. \autoref{tab:sota-comparison} shows how all considered approaches exhibit decreased performance. \name proves to be the best approach with an F1-score of $91.03\%$, followed once again by SGBot, with an F1-score of $88.19\%$. The remaining approaches struggle considerably on this dataset, trailing the leading techniques by almost $10$ percentage points. Similar to Cresci--17, BotRGCN and RGT cannot be evaluated on this dataset due to missing relationship data. We also observe that results presented in \autoref{tab:our-split-results} and \autoref{tab:sota-comparison} differ significantly. This discrepancy arises because the results in \autoref{tab:our-split-results} include a larger number of users compared to the split provided by the authors in~\cite{feng2022twibot}.

\textbf{Twibot--22.}
On the Twibot--22 dataset, \name achieves the highest overall F1 score among all considered approaches, including graph-based DL models. Of the remaining approaches, the random forest proposed in Kudugunta~\cite{kudugunta2018deep} performs best, losing $\sim2$ percentage points of F1 compared to \name. BotRGCN and RGT show very good precision and poor recall, indicating that they struggle to consistently identify bot accounts. Finally, both SGBot and Efthimion~\cite{efthimion2018supervised} show considerably worse performance on this dataset.

\textbf{Fox--8.}
The Fox--8 dataset lacks information on user relationships, preventing the evaluation of graph-based models such as BotRGCN and RGT. As shown \autoref{tab:sota-comparison}, all approaches perform consistently well on this dataset, with the exception of~\cite{efthimion2018supervised}, which trails by $\sim10$ percentage points. However, \name outperforms the considered s.o.t.a. approaches across all metrics, further emphasizing the effectiveness of its design based on local sensitivity.

\textbf{Takeaway.} \name shows superior performance compared to the considered approaches in terms of F1 across all datasets, with the exception of Cresci--15 where it is outperformed by SGBot. We find these results remarkable, as all considered state-of-the-art techniques leverage ML or DL algorithms. Despite its simpler design, the combination of approximate nearest neighbor matching and local sensitivity enables \name to surpass existing approaches.

\subsection{Resource Usage Comparison}
\label{sec:resource_usage}
This section evaluates the resource usage of \name and compares it against other state-of-the-art approaches. For each approach, we report the preprocessing and training time in seconds, as well as memory usage in megabytes. \autoref{tab:timings-table} presents our results. We observe that \name is generally very fast and lightweight:
on Twibot-22, the largest dataset, it takes only 9.63 seconds to generate the MD-DNA encoding for each accounting and 10.48 seconds to prepare the hashed ground truth. Furthermore, these results were achieved with unoptimized Python code. The time required by \name can be reduced further by exploiting GPU parallelization for hash computation and classification, and through improved preprocessing parallelization. Memory requirements for Bothash are similarly low, using a maximum of 2GB of RAM during the preprocessing of the Twibot--22 dataset. These results highlight the scalability of \name and the capability of being deployed on less powerful machines.

Traditional machine learning-based approaches are typically fairly lightweight. Efthimion and SGBot show preprocessing time that is generally in line or slightly higher than \name. Training time is higher for both approaches, with Efthimion being almost one order of magnitude higher. Kudugunta's random forest approach requires negligible time for preprocessing, while training time is in line with \name. Memory usage is variable among ML-based approaches, with SGBot requiring very little memory (624MB) on Twibot--22, increasing up to almost 5GB for Efthimion. Finally, DL-based approaches BotRGCN and RGT require extensive computational resources for preprocessing and training. Both approaches require considerable preprocessing time to generate the graph structure necessary for training (27,900 seconds for Twibot--22, or $\sim 8$ hours). The training time, however, varies considerably between the two. BotRGCN shows the lowest training time among all considered approaches on Twibot--22. While this might seem surprising, the network used by BotRGN and the graph data are both small, and each epoch can run all training data in parallel in a single operation~\cite{feng2022twibot}, which is extremely efficient. On the other hand, RGT uses a more complex model and graph data, resulting in much longer training times: 1656 seconds on Twibot--22. Similar behavior is also noted in the memory usage of the two approaches, with BotRGCN using 1.4GB and RGT requiring 22GB of memory.

Overall, \name demonstrates lower resource usage compared to all evaluated approaches, with the exception of Kudugunta's random forest, where its resource usage is comparable. Nevertheless, \name achieves significantly better performance across all tasks, as outlined in Section~\ref{sec:sota_comparison}.

\subsection{Generalization, Early Detection, and Ground Truth Size Analysis}
\label{subsec:generalization_edetection}
This section evaluates the generalization capabilities of \name across different datasets, its ability to detect bots when few posts are available for each user, and its performance with limited ground truth data.

\begin{table*}[]
\centering
\caption{Results of the generalization analysis. This table contains the best experiments performed by \name using a dataset as ground truth and another one as test set. We also reported the hyperparameters used to obtain these results.}
\label{tab:cross-dataset-table}
\resizebox{0.9\textwidth}{!}{%
\begin{tabular}{@{}cccccclccc@{}}
\toprule
\textbf{Ground Truth} & \textbf{Test set} & \multicolumn{4}{c}{\textbf{Metrics}} &  & \multicolumn{3}{c}{\textbf{Parameters}} \\ \midrule
 &  & \textbf{F1-score} & \textbf{Accuracy} & \textbf{Precision} & \textbf{Recall} &  & \textbf{K-shingle} & \textbf{Threshold} & \textbf{Alphabets} \\ \cmidrule(lr){3-6} \cmidrule(l){8-10} 
Cresci--15 & Cresci--17 & 98.27 & 96.94 & 99.08 & 97.47 &  & 2 & 0.2 & \btype + \btemporal \\
Cresci--15 & Cresci--18 & 98.56 & 97.37 & 97.74 & 99.40 &  & 15 & 0.9 & \btype \\
Cresci--15 & Twibot--22 & 53.19 & 56.79 & 52.59 & 53.81 &  & 3 & 0.1 & \btype + \bcontent + \btemporal \\
Cresci--15 & Fox--8 & 76.62 & 74.83 & 71.62 & 82.36 &  & 2 & 0.1 & \btemporal \\ \midrule
Cresci--17 & Cresci--18 & 98.55 & 97.35 & 97.71 & 99.40 &  & 15 & 0.9 & \btype + \bcontent + \btemporal \\
Cresci--17 & Twibot--22 & 45.94 & 59.32 & 58.34 & 37.88 &  & 3 & 0.1 & \bcontent \\
Cresci--17 & Fox--8 & 64.78 & 69.78 & 77.92 & 55.43 &  & 3 & 0.2 & \btype + \bcontent + \btemporal \\ \midrule
Cresci--18 & Twibot--22 & 54.13 & 41.89 & 74.55 & 42.49 &  & 2 & 0.4 & \btype \\
Cresci--18 & Fox--8 & 77.57 & 72.41 & 65.42 & 95.26 &  & 2 & 0.8 & \bcontent \\ \midrule
Twibot--22 & Fox--8 & 62.28 & 60.43 & 59.56 & 65.26 &  & 2 & 0.6 & \btype + \bcontent + \btemporal \\ \bottomrule
\end{tabular}%
}
\end{table*}
\begin{table*}[]
\centering
\caption{F1-score and accuracy achieved by \name on the \quotes{All} dataset for each alphabet.}
\label{tab:all_results_table}
\resizebox{\textwidth}{!}{%
\begin{tabular}{@{}ccccccccccccccccccccc@{}}
\toprule
\textbf{Dataset} & \multicolumn{20}{c}{\textbf{Alphabets}} \\ \midrule
 & \multicolumn{2}{c}{\btype} &  & \multicolumn{2}{c}{\bcontent} &  & \multicolumn{2}{c}{\btemporal} &  & \multicolumn{2}{c}{\btype + \bcontent} &  & \multicolumn{2}{c}{\btype + \btemporal} &  & \multicolumn{2}{c}{\bcontent + \btemporal} &  & \multicolumn{2}{c}{\btype + \bcontent + \btemporal} \\ \cmidrule(lr){2-3} \cmidrule(lr){5-6} \cmidrule(lr){8-9} \cmidrule(lr){11-12} \cmidrule(lr){14-15} \cmidrule(lr){17-18} \cmidrule(l){20-21} 
 & F1 & Acc. &  & F1 & Acc. &  & F1 & Acc. &  & F1 & Acc. &  & F1 & Acc. &  & F1 & Acc. &  & F1 & Acc. \\ \cmidrule(lr){2-3} \cmidrule(lr){5-6} \cmidrule(lr){8-9} \cmidrule(lr){11-12} \cmidrule(lr){14-15} \cmidrule(lr){17-18} \cmidrule(l){20-21} 
\textbf{All} & 66.25 & 60.23 &  & 64.66 & 60.89 &  & 66.98 & 63.48 &  & 67.80 & 64.91 &  & \textbf{69.27} & \textbf{66.40} &  & 68.81 & 63.76 &  & 69.45 & 65.60 \\ \bottomrule
\end{tabular}%
}
\end{table*}

\subsubsection{Generalization}
\label{subsubsec:generalization}
We assess the generalization capabilities of \name through cross-dataset testing, where one dataset is used as the ground truth and another as the test set. We also ensure that temporal ordering across datasets is maintained. For example, Cresci-15 is used as the ground truth while Cresci-17 serves as the test set, but the reverse configuration is deliberately avoided. This is done to evaluate the effectiveness of using older bot data as ground truth for detecting more recent bots. In this experiment, we use as alphabet the best combination identified in \autoref{tab:our-split-results}. Furthermore, we also provide performance results of \name across the \emph{All} meta-dataset, which is obtained by merging all considered datasets together. This presents a significant challenge, as it includes many different types of bot accounts.

\textbf{Cross-Dataset.} \autoref{tab:cross-dataset-table} illustrates the best results achieved in the cross-dataset setting. We observe that using Cresci--15 as the ground truth, \name performs consistently well on the Cresci--17 and Cresci--18 datasets with performance that is comparable to the non-cross dataset evaluation. However, the performance decreases significantly for Twibot--22 and Fox--8. On the former, \name achieves an F1-score of $53.19\%$, compared to $64.25\%$ in \autoref{tab:our-split-results} and $75.83\%$ in \autoref{tab:sota-comparison}. Similarly, for Fox--8, the F1-score declines to $76.62\%$ from $98.39\%$. A similar trend is noted when using Cresci--17 and Cresci--18 as ground truth. This performance degradation can be attributed to the evolving and increasingly complex behavior of bot accounts, which pose greater challenges for detectors. Moreover, performance is also likely influenced by the different methods of dataset creation employed by the authors of the different datasets. The Cresci datasets generally include bot accounts that were directly verified as being bots. For instance, Cresci--15 comprises bots that were directly purchased for the study. Cresci--17 includes bot accounts that were publicly used for specific automated promotions. On the other hand, the Twibot-22 dataset is constructed based on an initial expert annotation of 1000 accounts, with the remaining accounts labeled by extrapolating from this established ground truth. Finally, as mentioned in Section~\ref{subsec:datasets}, Fox--8 comprises only recent LLM-powered social bots, which strongly differ from classical social bots. Overall, the cross-dataset generalization of \name is effective for datasets that are temporally close and exhibit somewhat similar behavioral characteristics among the bots. However, when ground truth and test set are temporally distant or involve entirely different types of behaviors (e.g., Twibot--22 and Fox--8), detection performance declines. This outcome is expected, as \name relies on approximate nearest neighbor matching for classification. Consequently, when bot behaviors evolve significantly over time (e.g., from 2017 to 2022) or when entirely new behaviors emerge, such as those exhibited by LLM-powered bots, it becomes essential to update the ground truth to maintain high classification performance. We emphasize that this challenge, referred to as \textit{temporal shift} in the ML field, is not unique to \name but is a common issue faced by many ML algorithms~\cite{yao2022wild}.

\textbf{All Meta-Dataset.} \autoref{tab:all_results_table} summarizes \name's performance on the All meta-dataset. As shown in \autoref{tab:datasets_table}, this dataset contains a total of $217,953$ accounts, which were randomly split in the ground truth and test set following a 70\%-30\% split. The best results are achieved using the MD-DNA encoding with the \btype and \btemporal alphabets, yielding an F1-score of $69.27\%$ and an accuracy of $66.4\%$. These results represent a significant improvement over single-encoding approaches, with the best-performing single-encoding method showing a reduction of $3$ percentage points in F1-score and $6$ percentage points in accuracy. This underscores the substantial impact of the MD-DNA technique and the incorporation of the temporal alphabet on improving classification performance. 

\subsubsection{Early Detection}
\label{subsubsec:early_detection}
We assess the early detection capabilities of \name by preprocessing all datasets to retain only the first $K$ tweets posted by each user. This evaluation provides a measure of how many posts \name requires to accurately classify an account, i.e., how quickly bots are detected after they start actively participating in social media.

\begin{figure}[t]
    \centering
    \includegraphics[width=\columnwidth]{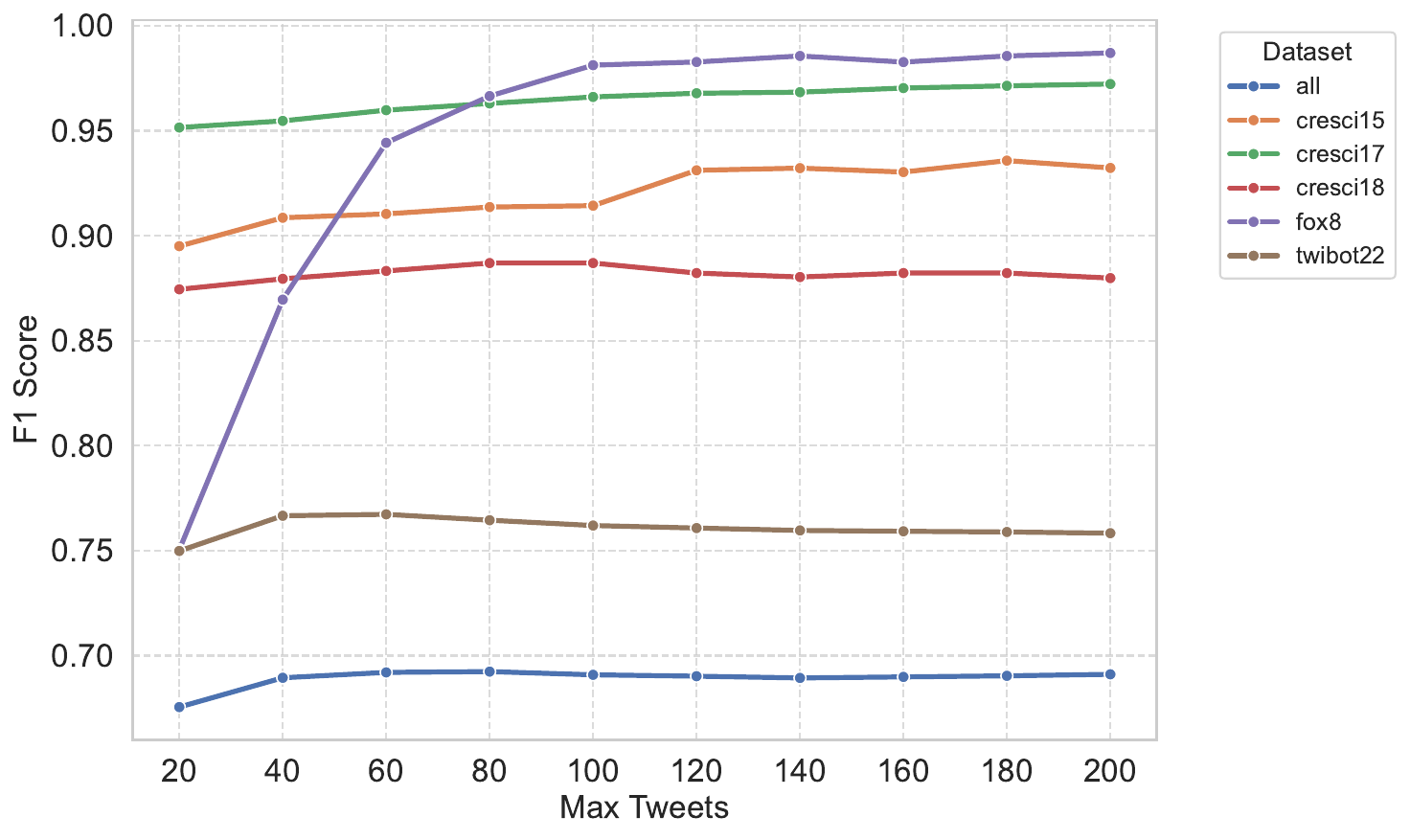}
    \caption{\name F1 performance across all considered datasets while limiting the number of available tweets for each account. Ranging from 20 to 200 tweets per account, with a step size of 20.}
    \label{fig:max_tweets}
\end{figure}

\autoref{fig:max_tweets} illustrates the F1-score achieved by \name for values of $K$ ranging from 20 up to 200 tweets. These experiments use the same ground truth - test set split provided by Feng et al.~\cite{feng2022twibot}, as presented in Section~\ref{sec:sota_comparison} and \autoref{tab:sota-comparison}. 
The results highlight that \name's performance is largely unaffected by the number of posts per user. In fact, with just $20$ tweets available, \name's performance is typically only a few percentage points lower than the optimal performance across all datasets. The only exception is the Fox--8 dataset, where the F1-score improves significantly as the number of tweets increases, plateauing at $\sim100$ tweets. It is challenging to identify the reasons for this discrepancy. We hypothesize that this behavior stems from the complex nature of LLM-powered bots, which are more adept at simulating human-like behaviors compared to traditional bots. As a result, more data is necessary to accurately detect subtle behavioral differences between LLM bots and real humans.
For the Cresci--15, Cresci--17, and Cresci--18 datasets, \name achieves an F1-score above $85\%$ with as few as $20$ tweets, and this performance remains relatively stable. A different trend is observed with the Twibot--22 dataset, where the highest F1-score ($76.66\%$) is obtained using $40-60$ tweets. Finally, for the All meta-dataset, \name demonstrates stable performance starting at $40$ tweets, with an F1-score of $68.94\%$, only marginally lower than the optimal performance of $69.27\%$ reported in \autoref{tab:all_results_table}.

\subsubsection{Ground Truth Size Analysis}
\label{subsubsec:gt_size_analysis}
For each dataset, we analyzed \name’s detection performance using varying portions of the data as ground truth. Specifically, in these experiments, we adjust the ground truth size between $10\% - 30\%$ of each dataset.

As shown in \autoref{fig:gt_size_analysis}, \name accurately classifies accounts across all datasets with as few as $10\%$ of the original data available. In the case of Cresci--15, \name achieves an F1-score of $93.54\%$ using a ground truth of $10\%$, which is $2$ percentage points lower than the optimal performance reached with $70\%$ ground truth. We observe minimal variation when increasing to $20\%$ and $30\%$. A similar pattern is observed for Cresci--17, where \name achieves an F1-score of $98.97\%$ using only $10\%$ of the ground truth data, which is comparable to the performance of $99.06\%$ reached with $70\%$ ground truth. In contrast, for Cresci--18, performance slightly decreases with $10\%$ ground truth to $87.71\%$, which is $\sim3$ percentage points off the optimal performance. For Fox--8 and Twibot--22, reducing the ground truth size incurs a somewhat higher performance penalty. On Fox--8, with a ground truth size of $10\%$ \name reaches $92\%$ F1, which improves to $97.14\%$ when the ground truth size is increased to $30\%$ --- only $\sim1$ percentage point lower than the optimal. Similarly, on Twibot--22, \name achieves a $70.97\%$ F1-score with a $10\%$ ground truth, which increases to $73.22\%$ at $30\%$. However, compared to the optimal result reported in \autoref{tab:sota-comparison}, the F1-score decreases by $\sim 2$ percentage points.
Finally, for the \quotes{All} dataset, \name achieves an F1-score of $68.15\%$ with $10\%$ ground truth size, which slightly improves to $68.95\%$ when the size is increased to $30\%$ --- comparable to the optimal results in \autoref{tab:sota-comparison}. 

Overall, \name's performance exhibits marginal correlation with the size of the ground truth. The performance remains largely stable across all datasets when the ground truth size is reduced from $70\%$ to $30\%$, with the largest drop observed on Twibot--22. Further reducing the ground truth size to $10\%$ leads to comparable performance on Cresci--15, Cresci--17, and All, while causing only a slight performance decline in the remaining dataset.
\section{Discussion and Limitation}
\label{sec:discussion}
This section examines the results presented in Section~\ref{sec:evaluation} and discusses the limitations of \name. 

\subsection{Results Discussion}
We structure the discussion of our results around the following research questions:

\begin{itemize}
    \item \textbf{RQ} \circled{1}: \textit{Is it possible to reliably detect bot accounts on OSN using non-machine learning methods?}

    \item \textbf{RQ} \circled{2}: \textit{Can the classification approach be generalized across different datasets?}

    \item \textbf{RQ} \circled{3}: \textit{Is early detection with limited account data feasible?}

    \item \textbf{RQ} \circled{4}: \textit{Are the resource savings achieved by \name significant compared to ML-based approaches?}

    \item \textbf{RQ} \circled{5}: \textit{Does MD-DNA represent an advancement over single D-DNA encodings?}
    
\end{itemize}

\begin{figure}[t]
    \centering
    \includegraphics[width=\columnwidth]{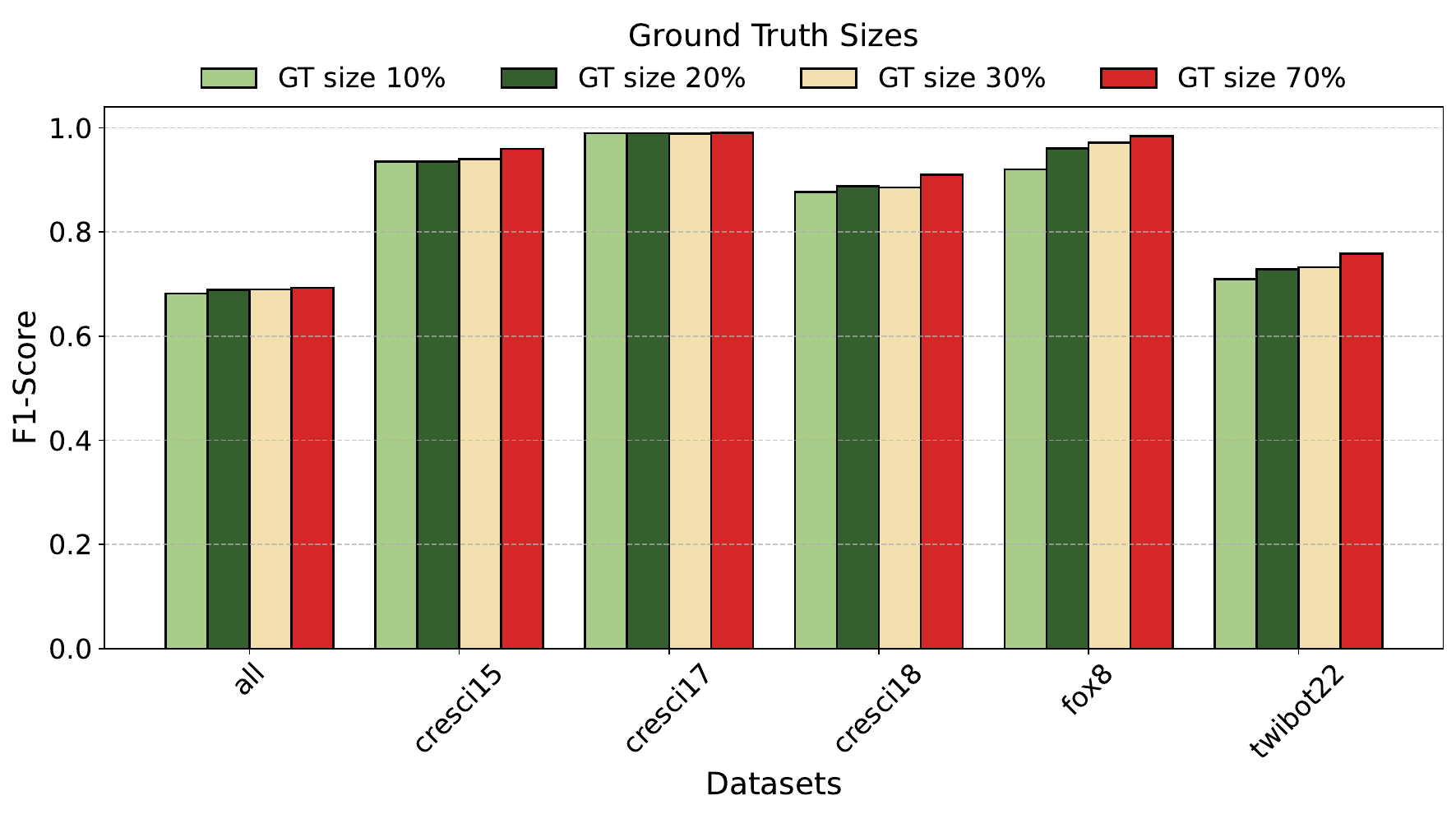}
    \caption{\name F1 performance across all considered datasets when using a limited portion of the ground truth (GT), ranging from $10\%$ to $30\%$. Comparison to the optimal results with $70\%$ ground truth size.}
    \label{fig:gt_size_analysis}
\end{figure}

We evaluated our approach across multiple datasets, demonstrating its effectiveness in detecting various types of social bots. Nonetheless, the performance of \name varies significantly across different datasets, reflecting the inherent challenges of bot detection in diverse social media contexts. Notably, \name achieved exceptional performance on the three Cresci datasets, despite the difficulties posed by Cresci--18, which focuses exclusively on specific tweet categories. Similarly, \name excelled on the LLM-based Fox--8 dataset, showcasing its robustness against complex social bots. Additionally, \name displays satisfactory performance on Twibot--22, which presents a higher classification challenge due to the inclusion of heterogeneous social bot types. Nonetheless, \name consistently outperforms a variety of state-of-the-art approaches that utilize diverse feature sets for classification across several datasets.
Therefore, our experimental results prove a positive answer to RQ \circled{1}.

Regarding the generalization capabilities of \name, our evaluation confirms that the proposed approach generalizes reasonably well, provided that bot behaviors do not change significantly, as shown in \autoref{tab:cross-dataset-table}. Indeed, the most significant discrepancies arise when using the Cresci datasets as ground truth and Twibot--22 as the test set. In this setting, the best result is achieved with Cresci--18, where \name achieves an F1-score of $54.13\%$, $\sim 20\%$ lower than the values reported in \autoref{tab:sota-comparison}. These results highlight the evolution of bot account behaviors on OSNs and the challenges associated with temporal shifts in datasets. Finally, we note that \name performs well on the All meta-dataset, achieving an F1-score of $69.27\%$. These results provide a partially positive answer to RQ \circled{2}; as long as the ground truth data does not diverge too much, in temporal or behavioral terms, \name generalizes well. However, when temporal shift in the data becomes significant, classification performance decreases markedly. We stress that this challenge is not unique to \name, but is a common problem with machine learning-based approaches~\cite{yao2022wild}.

Regarding early detection and performance and limited ground truth data, \autoref{fig:max_tweets} and \autoref{fig:gt_size_analysis} highlight some important findings. When simulating early detection by restricting the number of tweets per user, \name demonstrates promising results, with detection performance remaining largely stable across all datasets. The notable exception is Fox--8, where optimal performance is achieved with 100 tweets. We attribute this to the complex nature of LLM-powered bots, which are more adept at simulating human-like behaviors compared to traditional bots. Similarly, \name performs consistently well when limiting the ground truth data available for classification, with marginal performance drops moving from $70\%$ ground truth to $30\%$, or even $10\%$ in some datasets.
These findings prove a positive answer to RQ \circled{3}.

With respect to resource usage, \name is very efficient. A non-optimized implementation of our proposed approach requires only $\sim20$ seconds for preprocessing and ground truth generation for the largest dataset, Twibot--22. These results compare extremely favorably to DL-based approaches, which require over 8 hours for preprocessing and training when running on powerful hardware. Considering ML-based techniques, time requirements vary considerably between different approaches. In general, we see that \name outperforms all but one considered techniques. Memory usage across the different approaches varies more significantly. \name demonstrates moderately low memory usage, surpassing all but one of the considered approaches. It is important to note that our implementation of \name is \textit{not} optimized, unlike the other approaches that utilize well-optimized libraries. Our code runs on the CPU and is written entirely in pure Python. Resource usage can be further reduced by implementing core functions and data structures in more efficient languages such as C. Additionally, hash computations can be massively parallelized by leveraging GPU implementations. Furthermore, while ML and DL models typically require retraining when significant new training data becomes available, with \name, it is sufficient to process the new data and incorporate it into the ground truth, without the need to reprocess all previous data. This does not impact classification time, as the constant search time is ensured by our hash collision-based classification method.

Overall, our findings provide a partially positive answer to RQ \circled{4}, demonstrating significant temporal savings and average memory savings --- though with considerable margins for improvement.

\begin{table}[]
\centering
\caption{Hyperparameters used by \name on different dataset.}
\label{tab:hyperparameters_table}
\resizebox{0.8\columnwidth}{!}{%
\begin{tabular}{@{}cccc@{}}
\toprule
\textbf{Dataset} & \multicolumn{3}{c}{\textbf{Hyperparameters}} \\ \midrule
\textbf{} & \textbf{K-shingle} & \textbf{Threshold} & \textbf{Alphabet(s)} \\ \cmidrule(l){2-4} 
\textbf{Cresci--15} & 2 & 0.6 & \bcontent \\
\textbf{Cresci--17} & 4 & 0.4 & \btype \\
\textbf{Cresci--18} & 11 & 0.4 & \btype \\
\textbf{Twibot--22} & 4 & 0.1 & \bcontent + \btemporal \\
\textbf{Fox--8} & 7 & 0.3 & \btype + \btemporal \\ \bottomrule
\end{tabular}%
}
\end{table}

With regard to the advantages of the MD-DNA (RQ \circled{5}) approach compared to single D-DNA encodings, \autoref{tab:our-split-results} demonstrates that MD-DNA outperforms single D-DNA encodings across nearly all considered datasets. The performance improvement is particularly notable in recent datasets, such as Twibot--22 and especially Fox--8. On Fox--8, MD-DNA shows a four percentage points improvement over the best-performing single D-DNA encoding alphabet. Additionally, we highlight the contribution of the \btemporal alphabet, proposed in Section~\ref{sec:encoding}, to better classification performance on complex datasets like Twibot--22 and Fox--8. MD-DNA enables the encoding of more information into DNA sequences, allowing effective classification of more recent bot behaviors. Lastly, we emphasize that the results obtained with MD-DNA on the simpler Cresci datasets align with those achieved using the single D-DNA encoding.

\subsection{Limitations.}
Despite the advantages discussed, \name has some limitations. The first limitation arises from one of its strengths. We specifically designed \name to use a simplified user encoding based on DNA sequences to minimize resource requirements while maintaining good performance. While our evaluation demonstrates that this representation effectively captures meaningful features of user accounts, it also overlooks several other features that could be utilized to further enhance classification performance. Developing a new user encoding scheme that incorporates these additional features while maintaining a lightweight design presents an interesting direction for future research. Additionally, features derived from relationships (e.g., following/follower connections) were not considered by \name. Previous research has shown that these features provide rich contexts that can enhance social bot detection performance~\cite{feng2021twibot,feng2022twibot}. Incorporating such information into MD-DNA encodings is challenging, but it represents a promising avenue for further strengthening \name.

\section{Conclusions}
\label{sec:conclusions}
This study proposed \name, a novel, training-free approach to the social bot detection problem, which remains an open challenge. We demonstrated that \name is capable of accurately classifying bot accounts across various datasets. Our approach achieves significant results even when limited labeled data and account activity are available, enabling early detection of bot accounts. Additionally, the analysis of the All meta-dataset and cross-dataset scenarios indicates that \name generalizes effectively on unseen data, even when source and target domains differ.
As future work, we aim to integrate currently unused features into \name to further enhance its performance. Moreover, we plan to extend our approach to other OSN platforms such as Facebook, Instagram, and Bluesky.


\bibliographystyle{elsarticle-harv} 
\bibliography{sample-base}

\end{document}